\documentclass[%
preprint,
unsortedaddress,
 amsmath,amssymb,
 aps,
]{revtex4-2}

\usepackage{graphicx}
\usepackage{dcolumn}
\usepackage{bm}
\usepackage[locale=US,per-mode=fraction,separate-uncertainty,sticky-per,retain-explicit-plus]{siunitx}

\begin{document}


\title{Imaging Localized Variable Capacitance During Switching Processes in Silicon Diodes by Time-Resolved Electron Holography}

\author{Tolga Wagner}
 \affiliation{Technische Universit{\"a}t Berlin, Institute of Optics and Atomic Physics, Stra{\ss}e des 17. Juni 135, Berlin 10623, Germany}
 \email{tolga.wagner@physik.tu-berlin.de}

\author{H{\"u}seyin \c{C}elik}%
 \affiliation{Technische Universit{\"a}t Berlin, Institute of Optics and Atomic Physics, Stra{\ss}e des 17. Juni 135, Berlin 10623, Germany}

\author{Dirk Berger}%
 \affiliation{Technische Universit{\"a}t Berlin, Center for Electron Microscopy, Stra{\ss}e des 17. Juni 135, Berlin 10623, Germany}

\author{Ines H{\"a}usler}%
 \affiliation{Humboldt-Universit{\"a}t zu Berlin, Department of Physics, Unter den Linden 6, Berlin 12489, Germany}

\author{Michael Lehmann}%
 \affiliation{Technische Universit{\"a}t Berlin, Institute of Optics and Atomic Physics, Stra{\ss}e des 17. Juni 135, Berlin 10623, Germany}

\date{\today}

\begin{abstract}
Interference Gating or iGate is a unique method for ultrafast time-resolved electron holography in a transmission electron microscope enabling a spatiotemporal resolution in the $\si{\nm}$ and $\si{\ns}$~regime with a minimal technological effort. Here, iGate is used for the first image-based investigation of the local dynamics of the projected electric potential in the area of the space charge region of two different general purpose silicon diodes during switching between unbiased and reverse biased condition with a temporal resolution of $\SI{25}{\ns}$ at a repetition rate of $\SI{3}{\mega\Hz}$. The obtained results for a focus-ion-beam-prepared ultrafast UG1A rectifier diode, which shows a decreasing capacitance with increasing reverse bias are in good agreement with an electric characterization of the macroscopic device as well as with theoretical expectations. For a severely modified 1N4007 device, however, time-resolved electron holography revealed a MOSCAP-like behavior with a rising capacitance in the area of the space charge region during the switching into reverse biased condition. Remarkably, a different behavior, dominated by the effective capacitance of the electrical setup, can be observed in the vacuum region outside both devices within the same measurements, clearly showing the benefits of localized dynamic potentiometry.
\end{abstract}

\maketitle


\section{\label{sec:introduction}Introduction}

The semiconductor industry with its continuous developments plays a central role in our modern technology-driven society. Here, the theoretical understanding of underlying physical processes (e.g. from analytical models or numerical simulations) and experience-based knowledge in processing (e.g. from experimental observations) go hand in hand. This is particularly noticeable in the course of the continuous miniaturization of semiconductor structures (e.g. semiconductor nanostructures) in order to increase their efficiency while reducing their switching times \cite{moore1965}.

Probably the most established transmission electron microscopic (TEM) and thus imaging technique for the investigation of electrical potential distributions in semiconductor structures in the $\si{nm}$~range is off-axis electron holography (EH) \cite{Lehmann2002,Lichte2008}. By overlapping a reference wave with an object wave, an interference patterns can be acquired, which, via a numerical reconstruction process, grants access to the phase modulation $\varphi_{el}\left(x\right)$ of the reconstructed electron wave, which is directly proportional (by an interaction constant $\sigma$) to the projected potential distribution $V\left(x, y, z\right)$ inside and outside investigated samples \cite{Yazdi2015}:
\begin{equation}
    \label{eq:EH-phase-projection}
    \varphi_{el}\left(x, y\right) = \sigma \int V\left(x, y, z\right) \mathrm{d}z.
\end{equation}
In static experiments, EH has been able to demonstrate, among other things, the increasing influence of surface effects (e.g. crosstalk due to stray fields \cite{Razif2019} or the broadening of space charge regions (SCR) \cite{Yazdi2015,Cooper2009}) with a continuous reduction of the structure dimensions, thus serving as input for new theoretical models \cite{Nipane2017,Nipane2018}. For a long time, however, EH has, within the exposure times of common detectors (in the sub-second range), been limited to static investigations \cite{Houdellier2019}, so that observations of ultrafast dynamic effects (e.g. charge carrier transients or the effects of a dynamic variation of the capacitance, both of which contribute to changed switching times of the devices) have not been possible so far.

A novel method for realizing robust time-resolved experiments in the TEM, called interference gating (iGate) \cite{Niermann2017,Lehmann2020}, seems promising to address this shortcoming. By a deliberate destruction of the interference pattern by controlled external disturbances, synchronized to a periodic process with periodicity $T$ under investigation, a large part of the interferometric information is suppressed. Only for a fixed fraction $\tau$ (gate) of the period $T$, an undisturbed interference pattern can build up. The ‘‘partially disturbed’’ interference pattern acquired over an arbitrary number of periods represents a time-resolved electron hologram, with the numerical reconstruction process acting as a temporal filter that only retains the information from the gate. With this scheme, in principle, time resolutions into the picosecond ($\si{\ps}$) range can be realized \cite{Wagner2019}, making iGate a powerful tool for dynamic investigations of electronic nanostructures.

As a first real-world application of iGate, localized capacitive effects in periodically switched general-purpose Silicon (Si) diodes are investigated with $\si{\nm}$~spatial and $\si{\ns}$~temporal resolution, addressing the question of to what extent electrical properties of samples can be characterized using dynamic imaging. 

\section{\label{sec:Sample}Sample}

\begin{figure}
    \includegraphics[width=\columnwidth]{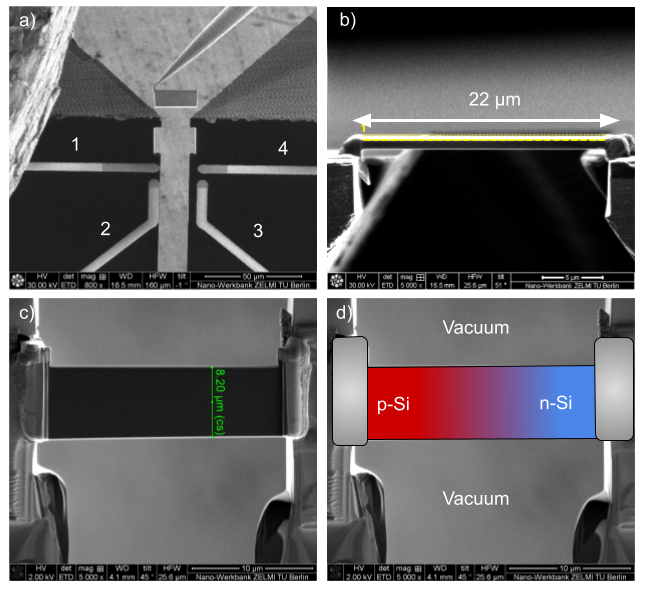}
    \caption{Contacting and post processing of the TEM lamella on a FIB optimized E-chip: a) conventional lamella transfer with the micromanipulator, where the numbers 1 to 4 denote the contact pads, b) thinning and polishing of the surfaces (view in cross-section), c) SEM image of the final contacted lamella with d) overlaid markings of the contacts, the p and n regions and the reference and object wave areas for the electron holographic acquisition.}
    \label{fig:TEM-lamella-specimen-prep}
\end{figure}

As objects of interest, two general-purpose Silicon diodes, a fast switching UG1A diode and a common 1N4007 rectifier diode, are investigated. Since both are manufactured by a thermal diffusion process, a strongly extended space charge region (SCR) in the $\si{\um}$~range is expected. The diodes are ground out of their housing (via a T-tool grinding device \cite{Zhang1998}) and the area of the p-n junction (located by voltage contrast microscopy \cite{Elliott2002,Nakamae1981}) is lifted out, transferred, contacted, thinned, and cleaned (fig.~\ref{fig:TEM-lamella-specimen-prep}) via focused ion beam (FIB) preparation (FEI Helios NanoLab 600, $\SI{30}{\kilo\volt}$ Ga-ion beams of $\SI{2.8}{\nano\ampere}$, $\SI{280}{\pico\ampere}$, and $\SI{93}{\pico\ampere}$) into a TEM lamella with dimensions of approx. $\SI{22}{\um} \times \SI{8}{\um} \times \SI{400}{\nm}$ (fig.~\ref{fig:TEM-lamella-specimen-prep}b, c). For the flat fixation and proper contacting of the TEM-lamella on the E-Chip, the thick transfer needle must lower the lamella without any physical interaction between the needle and the E-Chip. Therefore, a tall spacer lug between lamella and needle (fig.~\ref{fig:TEM-lamella-specimen-prep}a) is prepared with ion beam ($\SI{30}{\kilo\volt}$, $\SI{10}{\pico\ampere}$) induced platinum deposition prior to the mentioned lift-out process. The spacer lug is thereby slightly tilted with respect to the lamella to have enough mechanical freedom of movement. The lamellae are contacted at pads 1 and 4 in such a way that their interface of the p-n junction sits parallel between the free-standing electrodes (false colored in fig.~\ref{fig:TEM-lamella-specimen-prep}d) of the carrier chip (Protochips FIB optimized E-Chip, fig.~\ref{fig:TEM-lamella-specimen-prep}a). By this, parasitic currents (e.g. due to Platinum halo on the surface of the carrier chip) are avoided and the diode current is guided directly through the lamella (and its surface).

The preparation process was monitored by $I$-$V$-measurements (before and after FIB preparation), shown for both diodes in the Supplemental Material fig.~1 \cite{suppl-material}. After subtracting a shunt resistance (due to surface conduction), the FIB-prepared diodes exhibit a strongly non-linear behavior, which is similar to the expected diode behavior for the UG1A diode. The prepared 1N4007 diode, however, shows a different $I$-$V$-characteristic, featuring a saturation behavior for negative and positive biases. Nevertheless, both prepared diodes represent switchable devices (detailed description in the Supplemental Material \cite{suppl-material}). Furthermore, the switching times of both macroscopic devices (before preparation) were investigated (Supplemental Material fig.~2 \cite{suppl-material}) in small signal operation ($\SI{\pm 1}{\volt}$). For the switching into reverse bias condition, the macroscopic UG1A diode shows a switching time of approx. $t_{rr} = \SI{25}{\ns}$ (according to the data sheet, the expected switching time in full load operation is $t_{rr} = \SI{15}{\ns}$ \cite{UG1ADatasheet}). Under the same conditions, the macroscopic 1N4007 diode shows a switching time of approx. $t_{rr} = \SI{33}{\ns}$ (no information available in the data sheet \cite{1N4007Datasheet}).

\section{\label{sec:static-EH}Static Electron Holography}
\begin{figure}
    \includegraphics[width=\columnwidth]{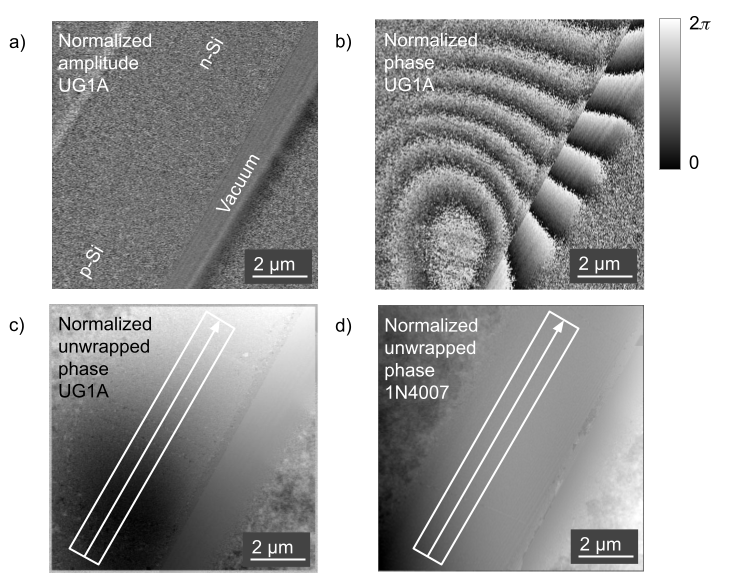}
    \caption{Reconstructed electron holograms of both diodes with a reverse bias of $U = \SI{-4}{\volt}$: a) normalized amplitude of the UG1A diode with markings for the p and n region as well as the vacuum region in front of the lamella, b, c) normalized wrapped (b) and unwrapped (c) phase of the UG1A diode, d) normalized unwrapped phase of the 1N4007 diode. Areas for the phase profiles (fig.~\ref{fig:UG1A_static_EH} and \ref{fig:1N4007_static_EH}) are indicated in the unwrapped phases.}
    \label{fig:rec-EH-pn}
\end{figure}

The FIB-prepared diodes were investigated with static electrical biasing EH. For this purpose, the FEI Titan 80-300 Berlin Special Holography TEM was used in a special Lorentz mode for wide fields of view \cite{Wagner2019} at an acceleration voltage of $U_a = \SI{200}{\kilo\volt}$. The electron holograms were acquired with an acquisition time of $t_{exp} = \SI{4}{\s}$ using a Gatan US1000 CCD. The vacuum region in front of the lamella (according to fig.~\ref{fig:TEM-lamella-specimen-prep}d) was utilized as reference wave area. Exemplary reconstructed holograms are shown in fig.~\ref{fig:rec-EH-pn}. For the static investigations, the electrical bias U is applied by a Keithley 2460 source meter, connected to a modified Protochips Aduro electrical biasing holder \cite{Wagner2022}.

As expected, the reconstructed amplitude (fig.~\ref{fig:rec-EH-pn}a) shows no modulation due to the potential distribution even with an applied reverse bias of $U = \SI{-4}{\volt}$, but the reconstructed phase clearly exhibits a modulation (fig.~\ref{fig:rec-EH-pn}b). The phase modulation includes not only the projected potential of the reverse biased sample itself, but also parasitic influences (stray field of the surface of the sample, stray field of the contacts, influences of the electron-optical beam path and unwanted modulation of the reference wave). The latter in particular lead to the unexpected bending of the equi-phase lines within the sample (fig.~\ref{fig:rec-EH-pn}b), which can be corrected by normalization. For the normalization, the reconstructed electron waves from measurements with applied bias are divided by reconstructed electron waves from measurements without applied bias. This compensates for all artifacts of the measurement setup (e.g. due to distortions or charges in the beam path), the sample (e.g. due to dynamic diffraction and charging) but also for the built-in potential. The normalized reconstruction thus only contains information resulting from a change of the applied bias.

\begin{figure}
    \includegraphics[width=\columnwidth]{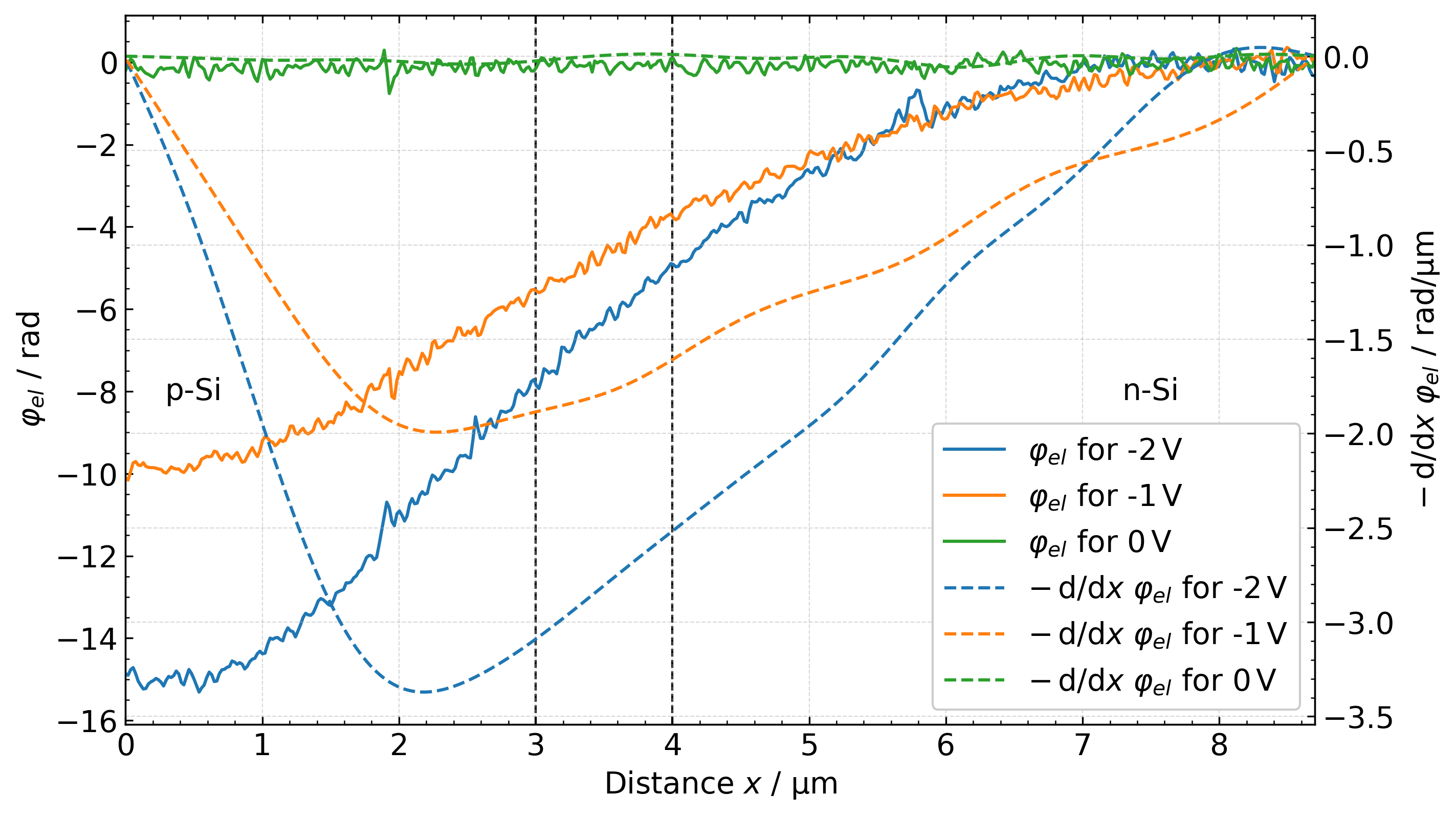}
    \caption{Plot of the normalized unwrapped phase profiles $\varphi_{el}\left(x\right)$ (phase offsetted to zero on the n-doped Side) as well as their first-order derivatives (phase slopes) $-\mathrm{d}/\mathrm{d}x~\varphi_{el}\left(x\right)$ (dashed lines) of the FIB-prepared UG1A diode in $x$-direction for different applied reverse biases U (taken from the area according to fig.~\ref{fig:rec-EH-pn}c) with the doped regions of the diode indicated near the edges of the plot. The vertical dashed lines mark the area for the evaluation of the dynamic phase slopes (fig.~\ref{fig:iGate_rectangle_UG1A} and \ref{fig:iGate_rectangle_1N4007}).}
    \label{fig:UG1A_static_EH}
\end{figure}

The profile plot of the normalized unwrapped phases $\varphi_{el}\left(x\right)$ of the UG1A diode (fig.~\ref{fig:UG1A_static_EH}) shows an unmodulated phase course for the unbiased condition and a non-linear phase course for applied reverse biases, whereas the phase jump (phase difference for the beginning and end of the phase profile) increases with applied reverse bias. Additionally, the phase slopes (first-order derivative $-\mathrm{d}/\mathrm{d}x~\varphi_{el}\left(x\right)$, which is proportional to the projected electric field) show a minimum around $x = \SI{2}{\um}$, which also increases with applied reverse bias. The derivatives were calculated by a priori smoothing of the phase profiles with a 1D~Gaussian Filter with $\sigma_G = \SI{100}{\nm}$.

\begin{figure}
    \includegraphics[width=\columnwidth]{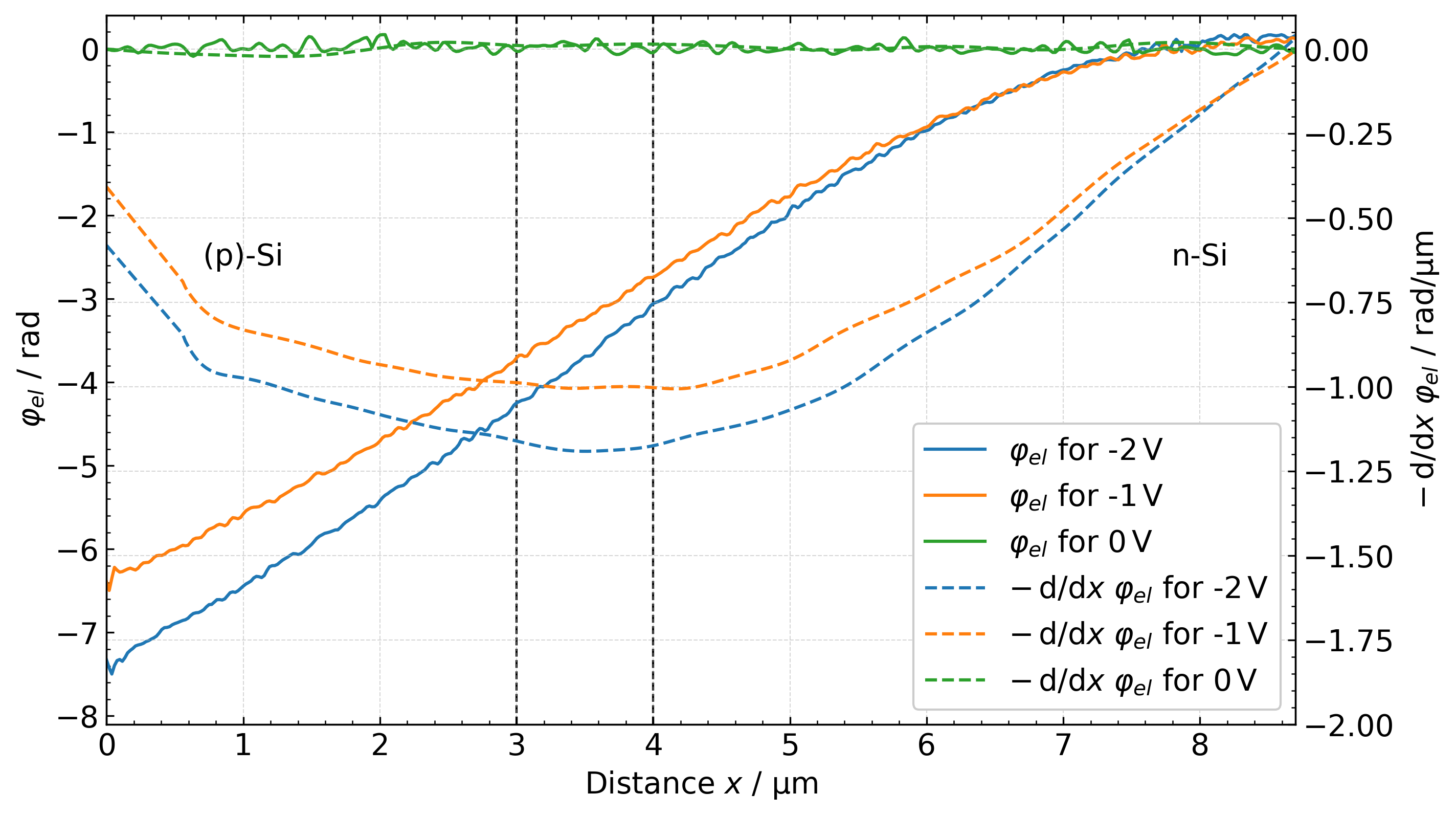}
    \caption{Plot of the normalized unwrapped phase profiles $\varphi_{el}\left(x\right)$ (phase offsetted to zero on the n-doped Side) as well as their first-order derivatives (phase slopes) $-\mathrm{d}/\mathrm{d}x~\varphi_{el}\left(x\right)$ (dashed lines) of the FIB-prepared 1N4007 diode in $x$-direction for different applied reverse biases $U$ (taken from the area according to fig.~\ref{fig:rec-EH-pn}d) with the doped regions of the diode indicated near the edges of the plot. The vertical dashed lines mark the area for the evaluation of the dynamic phase slopes (fig.~\ref{fig:iGate_rectangle_UG1A} and \ref{fig:iGate_rectangle_1N4007}).}
    \label{fig:1N4007_static_EH}
\end{figure}

Taking into account eq.~(\ref{eq:EH-phase-projection}), the reconstructed phases (fig.~\ref{fig:rec-EH-pn}) and their correlated profiles (fig.~\ref{fig:UG1A_static_EH} and \ref{fig:1N4007_static_EH}) are directly proportional to the projected electric potential distribution of the samples. For the UG1A diode, the phase profiles for the reverse biased condition (fig.~\ref{fig:UG1A_static_EH}) qualitatively agree with a projection of the typical potential profile of a p-n junction diode (e.g. quadratic approximation) with an asymmetric doping concentration. However, it must be taken into account that the electric stray field of the sample not only modulates the reference wave, but also leads to non-negligible phase modulations above and below the sample. These can, for example, result in an expansion of the width of the measured potential jump and thus hinder quantitative investigations. Nevertheless, the whole SCR appears to be inside the prepared TEM lamella for the UG1A diode, suggesting a successful FIB preparation of an intact diode.

The plot for the 1N4007 diode (fig.~\ref{fig:1N4007_static_EH}) paints a different picture. While the phases $\varphi_{el}\left(x\right)$ show a similar course towards the end of the profile (n doped side), they seem to be truncated towards the p doped side (at the beginning of the profile) resulting in a TEM lamella which only features a partial SCR. This was caused during the cut-out procedure due to the fact that the junction width of the 1N4007 diode seemingly exceeds the overall length of the lamella, which is limited by the width of the slit in the E-Chip. Consequently, neither a phase jump can be detected nor does the phase slope $-\mathrm{d}/\mathrm{d}x~\varphi_{el}\left(x\right)$ show a pronounced minimum (dashed lines). Although, in contrast to the UG1A diode, both profiles show only limited changes with increasing reverse bias for the 1N4007 diode, a difference to the unbiased condition is still clearly visible (due to the normalization).

According to the course of the phase, this sample is more likely to represent a capacitor with a poorly chosen dielectric (lightly p-doped Si on the left electrode and gradually increasing n-doped Si on the right electrode), which is supported by the limited change of the phase jump (fig.~\ref{fig:1N4007_static_EH}) for different reverse biases. On this basis, and considering the $I$-$V$-characteristics (Supplemental Material fig.~1 \cite{suppl-material}), a severely modified device with a different electric switching behavior than that of a conventional diode is expected.
\section{\label{sec:time-resolved-EH}Time-Resolved Electron Holography}
\begin{figure}
    \includegraphics[width=\columnwidth]{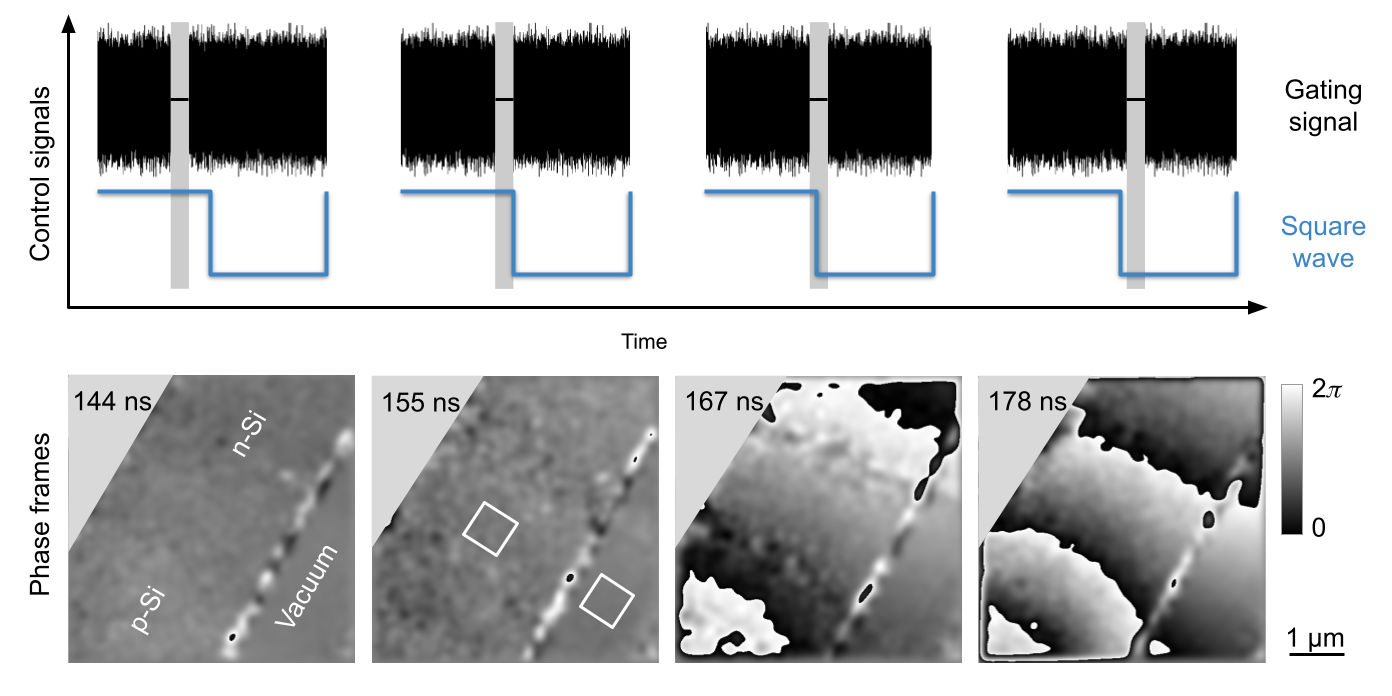}
    \caption{Schematic representation of the applied control signals during sampling and the corresponding reconstructed time-resolved wrapped phases (frames) with a temporal resolution of $\tau = \SI{25}{\ns}$ for the switching of the UG1A diode into reverse biased condition for different gate positions between $\SI{144}{\ns}$ and $\SI{178}{\ns}$ with a sampling resolution of $t_0 = \SI{11.1}{\ns}$.}
    \label{fig:frames}
\end{figure}

For the investigation of the switching dynamics of both diodes, interference gating \cite{Niermann2017} is used in order to produce time resolved electron holograms. For this, a GW Instek MFG-2260MRA multi-channel arbitrary signal generator producing synchronized control signals (a gating signal and a signal for the switching the diodes) was connected to a biprism in the condenser aperture plane (used as a gating device) and the electrical biasing holder (detailed description of the setup in \cite{Wagner2019}). Both signals, provided by a custom Python script, were monitored by an Oscilloscope (GW Instek GDS-1102B connected in parallel). For the switching of the diodes, a square wave with a peak-to-peak amplitude of $U = \SI{1.0}{\volt}_{pp}$ (from $U = \SI{0}{\volt}$ to $U = \SI{-1}{\volt}$) at a repetition rate of $f = \SI{3}{\mega\Hz}$ was applied. By this, the diodes are unbiased and reverse biased for half of the period $T = \SI{333.3}{\ns}$ respectively. For the gating, a noise signal with a peak-to-peak amplitude of $U = \SI{2.7}{\volt}_{pp}$ featuring a $U = \SI{0.0}{\volt}_{pp}$ gate with a gate length of $\tau = \SI{25}{\ns}$ was used to sample the switching process with a sampling resolution of $t_0 = \SI{11.1}{\ns}$, resulting in 33 phase frames (schematically shown in fig.~\ref{fig:frames}). The time-resolved electron holograms were acquired in a setup similar to the static ones with a slightly higher magnification (in order to increase the fringe contrast) in a central region of the diodes (around $x = \SI{4}{\um}$ in fig.~\ref{fig:UG1A_static_EH} and \ref{fig:1N4007_static_EH}). Since a major advantage of interference gating is that the time resolution is independent of the used detector, the time-resolved holograms, with an acquisition time of $t_{exp} = \SI{8}{\s}$ each, were likewise recorded with a Gatan US1000 CCD. The interference gating thus reduces the fringe contrast of the time-resolved holograms by the ratio between gate length $\tau$ and period $T$ (the so-called gate fraction) compared to a static hologram, leading to an increased phase noise \cite{Niermann2017,Wagner2019}. In order to increase the signal-to-noise ratio and compensate for the increased phase noise (to a certain degree), five time-resolved holograms for each gate position (frame) were averaged during the reconstruction \cite{Niermann2014}.

Exemplary phase frames (wrapped) for the switching into reverse biased condition of the UG1A diode are shown in fig.~\ref{fig:frames}.

The development of the phase modulation (comparable to the static case in fig.~\ref{fig:rec-EH-pn}b) can be seen step by step in the exemplary phase frames (e.g. by the equi-phase lines moving closer together). In addition, slight double exposure electron holography (DEEH) artifacts \cite{Migunov2017,Niermann2017} are visible in the frames at $t = \SI{155}{\ns}$ and $t = \SI{167}{\ns}$ (surrounding the white square in the central region of the UG1A diode). However, their influence on the phase modulation is limited, therefore the time-resolved phase information can be regarded as an extension to eq.~(\ref{eq:EH-phase-projection}) by means of a dynamic projected potential distribution.

\begin{figure}
    \includegraphics[width=\columnwidth]{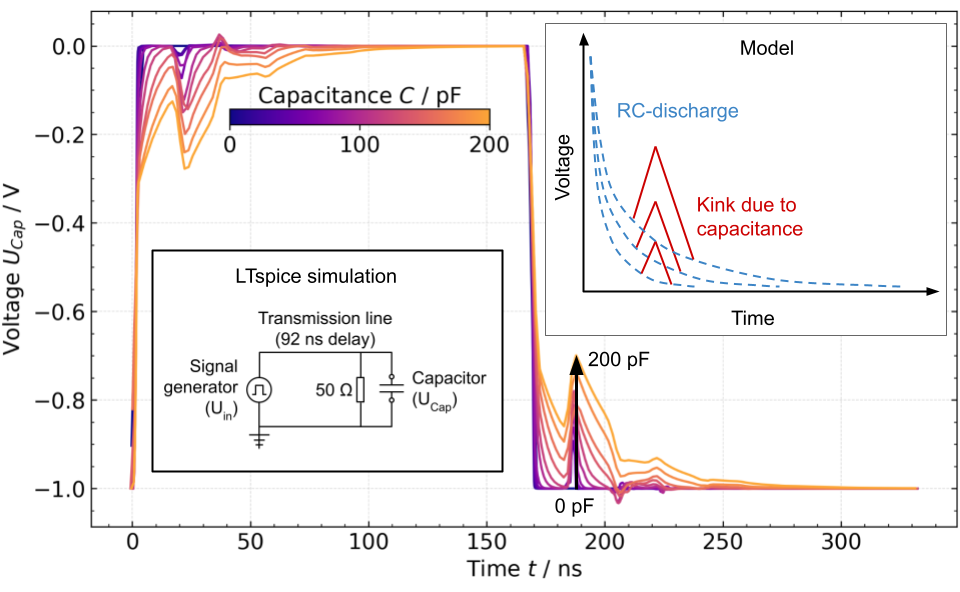}
    \caption{Plot of simulated dynamic voltages $U_{cap}\left(t\right)$ at a capacitor, driven by a $\SI{3}{\mega\Hz}$ square wave ($U_{in} = \SI{1}{\volt}_{pp}$) for different capacitances from $C = \SI{0}{\pico\farad}$ to $C = \SI{200}{\pico\farad}$ (circuit for the LTspice simulation embedded in the lower left box). Basic model for the illustration of the effect with varying capacitance (resulting kink due to reflections) embedded in the upper right box.}
    \label{fig:LTSpice-simulation}
\end{figure}

Considering the high frequency of $f = \SI{3}{\mega\Hz}$ of the applied square wave signal, capacitive effects of the overall system (cabling, electrical biasing holder, carrier chip and the diodes) play a non-negligible role.  In order to take these into account, a capacitive system (representative of the overall system), driven by a $\SI{3}{\mega\Hz}$ square wave signal ($U_{in} = \SI{1}{\volt}_{pp}$), was modeled by numerical simulation using LTspice \cite{LTSpice} (lower left box in fig.~\ref{fig:LTSpice-simulation}).

In the simulation, the capacitance of the system was varied from $C = \SI{0}{\pico\farad}$ to $C = \SI{200}{\pico\farad}$ for an effective transmission line delay of $\SI{92}{\ns}$ (determined by time domain reflectometry), yielding dynamic voltages $U_{cap}$ plotted in fig.~\ref{fig:LTSpice-simulation}. Without a capacitance ($C = \SI{0}{\pico\farad}$), $U_{cap}$ matches the applied square wave $U_{in}$, with increasing $C$, however, a kink starts building near the transition between the two voltages. The kink is primarily caused by signal reflections on the capacitor traveling back the transmission line \cite{carr2012practical}. The behavior of $U_{cap}$ can be described, to a first approximation, by a basic $R$-$C$ charge/discharge superimposed with a kink of variable height (upper right box in fig.~\ref{fig:LTSpice-simulation}). The height of the kink in this basic model is proportional to the capacitance $C$.

\begin{figure}
    \includegraphics[width=\columnwidth]{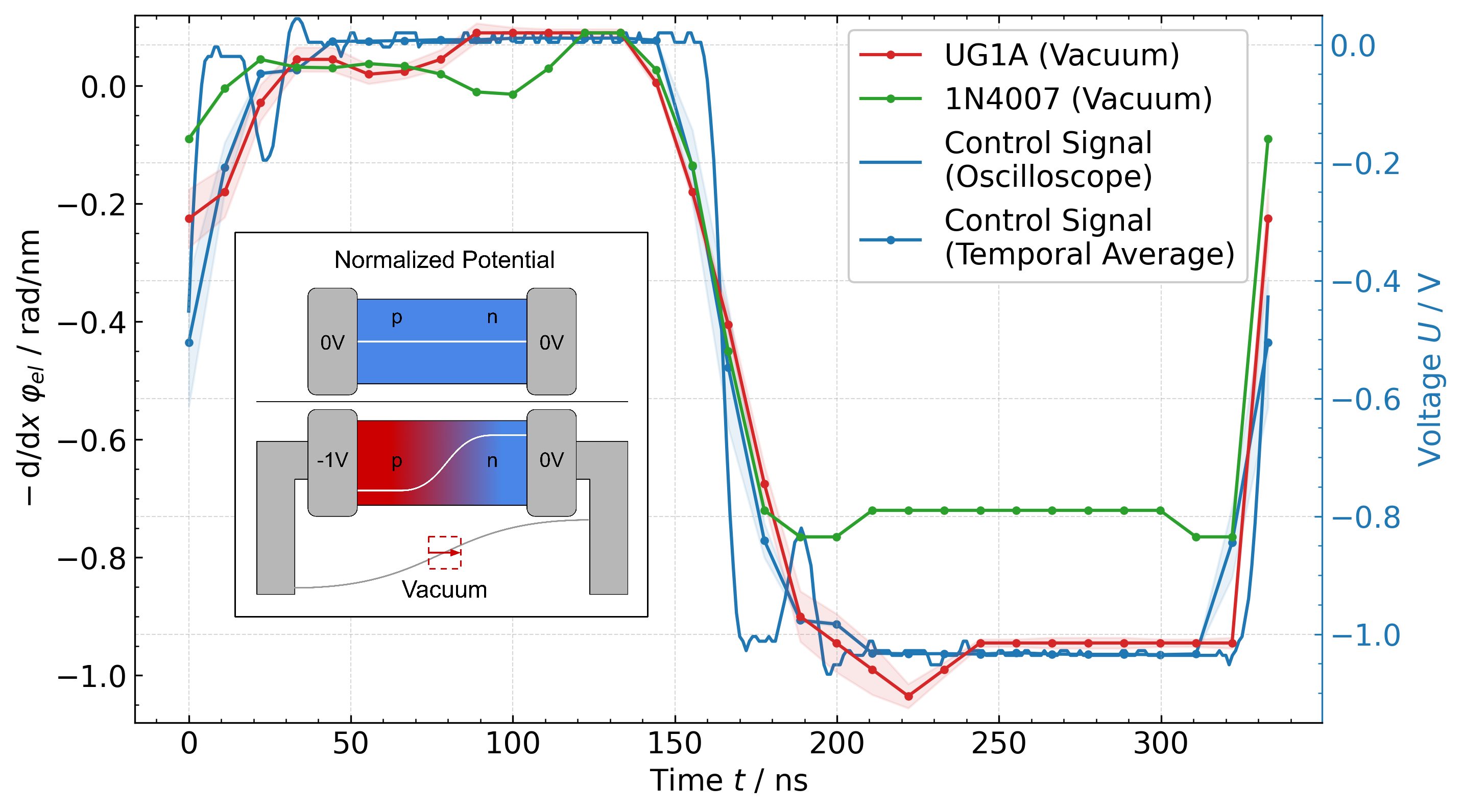}
    \caption{Plot of the applied control signal, measured by the oscilloscope (blue line), its temporal average (blue dots), calculated phase slopes of the phase frames, taken in the vacuum region in front of the lamella, for the UG1A diode (red dots) and the 1N4007 diode (green dots). The standard deviation (light colored) is given for all averaged signals. A schematic representation of the measurement is depicted in the lower left box.}
    \label{fig:iGate_vacuum_UG1A_1N4007}
\end{figure}

In the experiment, the applied control signal for the time-resolved electron holographic investigations, monitored by an oscilloscope (blue line in fig.~\ref{fig:iGate_vacuum_UG1A_1N4007}) and measured in the vacuum region in front of the TEM lamella, features a kink (similar to the simulated signals) as well.  Interestingly, the course of the monitored signal (e.g. the height of the kink) does not change for the different diodes (or even an empty carrier chip without any TEM lamella). This behavior is predominantly caused by the electrical biasing system itself (mainly the carrier chip). A TEM lamella, given by its relatively small dimensions (in contrast to the carrier chip), does not have a notable effect on the monitored signal at all (at least in this experimental setup of the TEM lamella, which can be regarded as a parallel setup of capacitors). In regard to the simulated signal (comparison of the height of the capacitive kinks), the overall capacitance of the biasing system can be determined to approx. $C = \SI{150}{\pico\farad}$.

Since in the time-resolved measurements, the dynamic phase modulation is averaged over the gate length of $\tau = \SI{25}{\ns}$ with a sampling resolution of $t_0 = \SI{11.1}{\ns}$, the monitored signal is also temporally averaged in the same manner ($\SI{25}{\ns}$ every $\SI{11.1}{\ns}$; blue dots in fig.~\ref{fig:iGate_vacuum_UG1A_1N4007}) in order to be comparable. By this, the transient between the two switching states broadens due to the averaging (comparable to a convolution with a rectangular pulse).

According to eq.~(\ref{eq:EH-phase-projection}), the phase frames themselves are proportional to the dynamic projected electric potential distribution $\varphi_{el}\left(x\right)$, therefore the phase slopes $-\mathrm{d}/\mathrm{d}x~\varphi_{el}\left(x\right)$ are proportional to the projected electric field strength distribution and thus correlated to an electric voltage distribution. Consequently, the phase slopes in the vacuum region in front of the lamellae (indicated by the white square in fig.~\ref{fig:frames}, plotted as red dots for UG1A diode and green dots for 1N4007 diode in fig.~\ref{fig:iGate_vacuum_UG1A_1N4007}), which are mainly modulated by the electric potential of the free-standing electrodes of the carrier chip, can be compared to the temporal averaged control signal shown as blue line in fig.~\ref{fig:iGate_vacuum_UG1A_1N4007}.

The course of the dynamic phase slopes $-\mathrm{d}/\mathrm{d}x~\varphi_{el}\left(x\right)$ in the vacuum region in front of the UG1A diode (red dots in fig.~\ref{fig:iGate_vacuum_UG1A_1N4007}) is in excellent agreement with the control signal. This is also true for the course of the dynamic phase slopes in the vacuum region of the 1N4007 diode (green dots in fig.~\ref{fig:iGate_vacuum_UG1A_1N4007}), whose scaling, however, deviates from that of the control signal. This is mainly caused by a larger distance between the free-standing electrodes of the carrier chip (manufacturing-related) for this lamella, which leads to a reduced field strength (for the same applied voltage $U$).

\begin{figure}
    \includegraphics[width=\columnwidth]{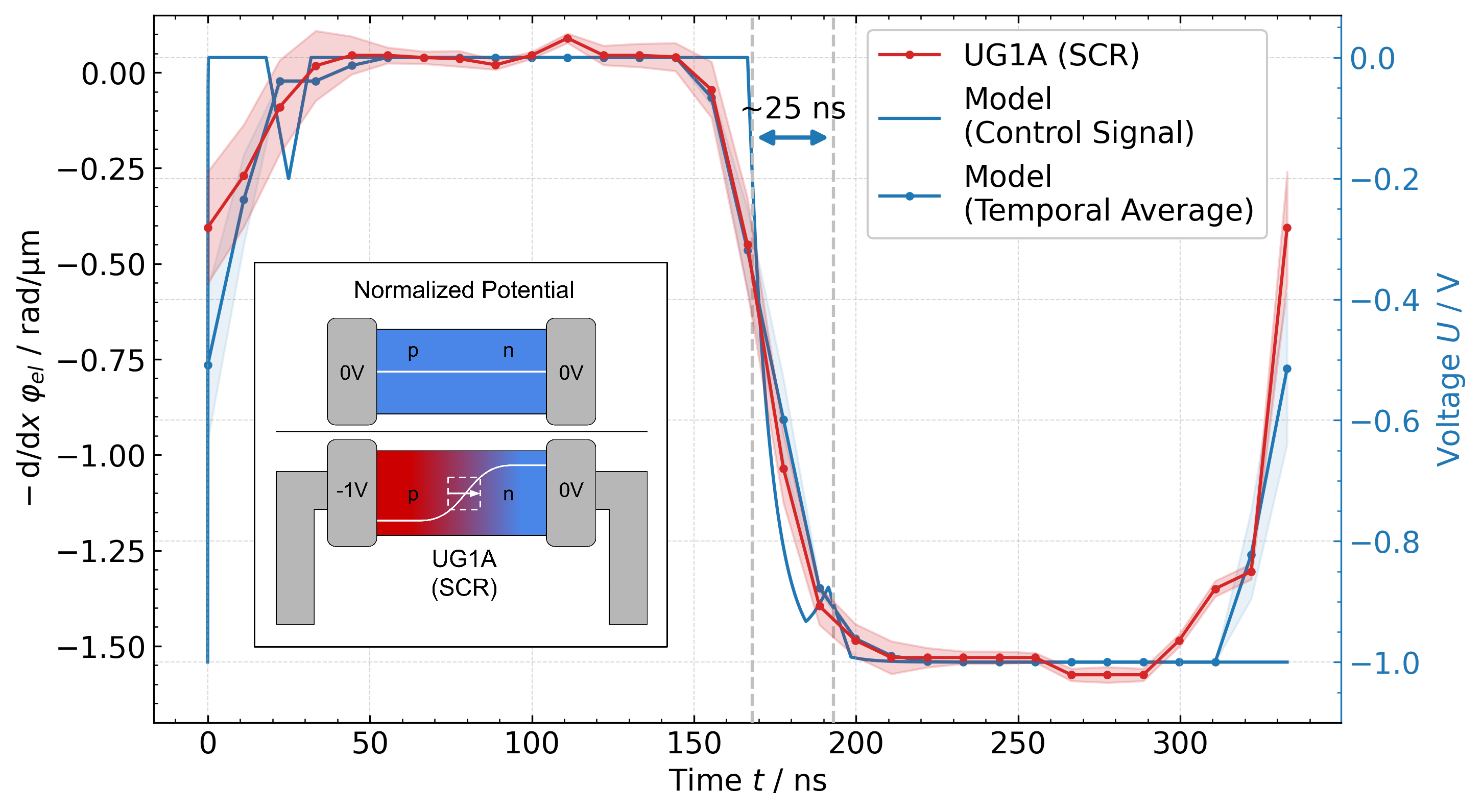}
    \caption{Plot of the best fitting modeled control signal, according to the basic model in the upper right box in fig.~\ref{fig:LTSpice-simulation} (blue line), its temporal average (blue dots), calculated phase slopes of the phase frames, taken in the center region of the lamella, for the UG1A diode (red dots). The standard deviation (light colored) is given for all averaged signals. A schematic representation of the measurement is depicted in the lower left box.}
    \label{fig:iGate_rectangle_UG1A}
\end{figure}

In contrast to the measurements in the vacuum region, the dynamic phase slopes in a central region of the UG1A diode (indicated by a white square in fig.~\ref{fig:frames}, plotted as red dots in fig.~\ref{fig:iGate_rectangle_UG1A}; notably within the same data set, only at a different spatial location) show a different course, primarily as a narrowing of the transient during switching into reverse bias condition.

This becomes particularly clear in comparison to the temporal average (blue dots) of the modeled control signal (blue line) in fig.~\ref{fig:iGate_rectangle_UG1A}. Here, a switching time of $t_{rr} = \SI{25}{\ns}$ (taken from the electrical investigation shown in the Supplemental Material fig.~2 \cite{suppl-material}) and capacitance of $C = \SI{7.5}{\pico\farad}$ was used for the basic model. For these values, the signals are in excellent agreement, yielding a capacitive kink that is significantly smaller compared to the measurements in the vacuum region (fig.~\ref{fig:iGate_vacuum_UG1A_1N4007}). This indicates a reduction of the capacitance of the diode during the transition into reverse biased condition. When switching back to the unbiased condition, the dynamic phase slopes in the diode region (fig.~\ref{fig:iGate_rectangle_UG1A}) behave almost identically to those in the vacuum region (fig.~\ref{fig:iGate_vacuum_UG1A_1N4007}), indicating an increase back to the initial capacitance. This behavior, called depletion capacitance \cite{sze2021physics}, is quite typical for diodes (fundamental principle of varicap diodes) and is primarily caused by a change in the width of the SCR due to an applied bias. Given by the small influence of the reduced capacitive kink, the transient time into reverse biased condition (according to the basic model) was determined to take approx. $\SI{25}{\ns}$, which is also in excellent agreement with the electrical characterization (Supplemental Material fig.~2 \cite{suppl-material}).

\begin{figure}
    \includegraphics[width=\columnwidth]{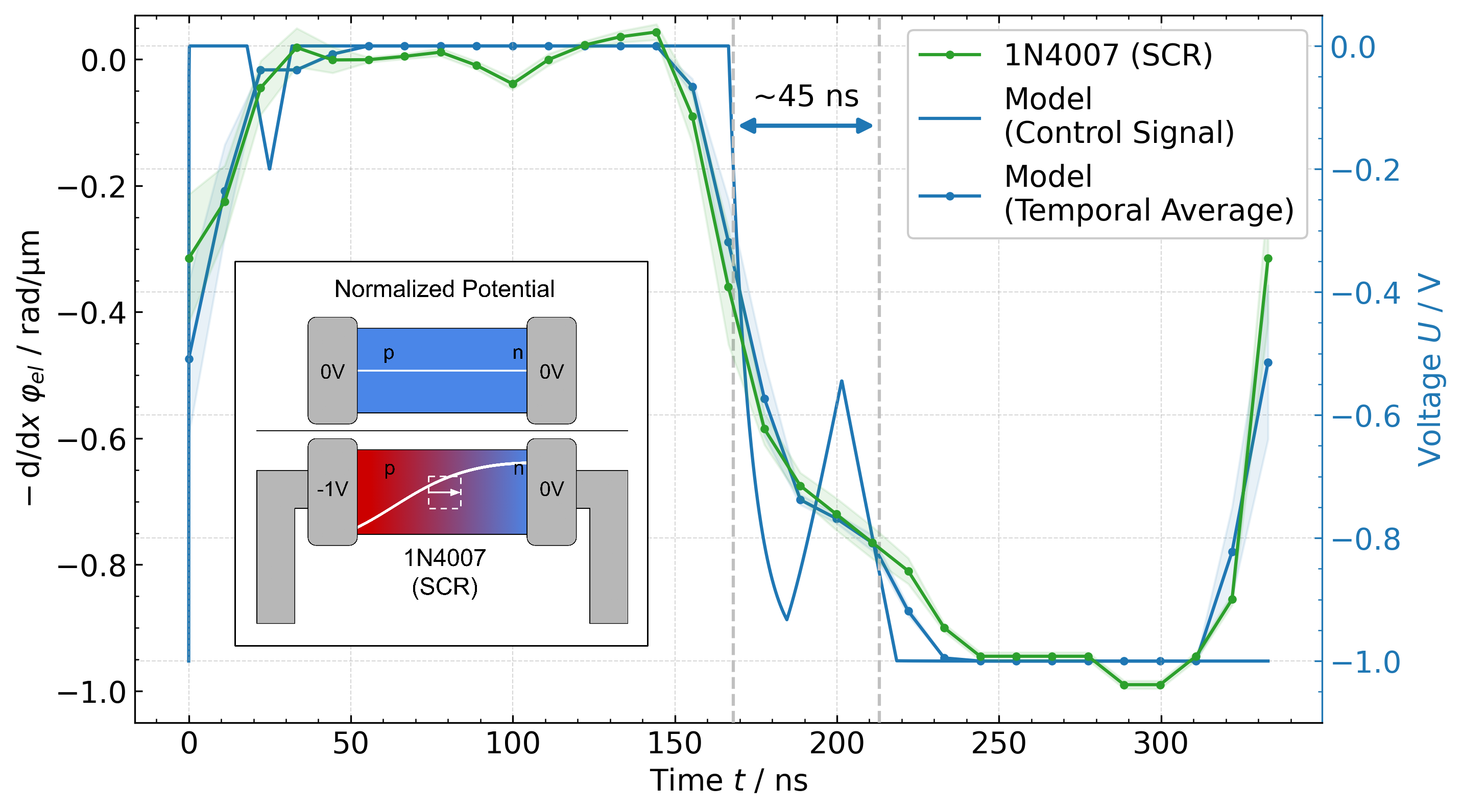}
    \caption{Plot of the best fitting modeled control signal, according to the basic model in the upper right box in fig.~\ref{fig:LTSpice-simulation} (blue line), its temporal average (blue dots), calculated phase slopes of the phase frames, taken in the center region of the lamella, for the 1N4007 diode (green dots). The standard deviation (light colored) is given for all averaged signals. A schematic representation of the measurement is depicted in the lower left box.}
    \label{fig:iGate_rectangle_1N4007}
\end{figure}

Even stronger deviations from the vacuum region are observed for the dynamic phase slopes in a center region of the 1N4007 diode (green dots in fig.\ref{fig:iGate_rectangle_1N4007}). For a nearly excellent agreement with the temporal average (blue dots) of the modeled control signal (blue line), a switching time of $t_{rr} = \SI{33}{\ns}$ (taken from the electrical investigation shown in the Supplemental Material fig.~2 \cite{suppl-material}) and a capacitance of $C = \SI{430}{\pico\farad}$ was used in the basic model. By this, the height of the capacitive kink is drastically increased compared to the vacuum region, resulting in an overall extended transition time of approx. $\SI{45}{\ns}$. Under the (aforementioned) assumption of a capacitor with lightly doped Si as dielectric, such a behavior could be related to a reduction of the free charge carriers in the doped silicon by the applied ‘‘reverse’’ bias. Such behavior is typical for metal-oxide-semiconductor capacitors (MOSCAP), where reverse bias leads to an accumulation of charge carriers at the interfaces and thus to a depletion zone inside the oxide, resulting in an increasing capacitance \cite{sze2021physics}. Similar to the UG1A diode, the transition back into the unbiased condition seems to be unchanged (compared to the measurement in the vacuum region in fig.~\ref{fig:iGate_vacuum_UG1A_1N4007}). 

\section{\label{sec:conclusion}Conclusion and Outlook}
It was shown that iGate is well suited for the investigation of dynamic electrical potential distributions and their effects in semiconductor devices.  Here, it was used to image the effects of localized dynamic capacitance in switching general purpose Silicon diodes with a temporal resolution of $\tau = \SI{25}{\ns}$. For a successfully prepared fast switching UG1A diode, a decreasing capacitance (down to approx. $C = \SI{7.5}{\pico\farad}$) during switching into reverse biased condition was observed inside the sample, whereas within the same measurement, the capacitance outside the sample remained unchanged at approx. $C = \SI{150}{\pico\farad}$. For a severely modified 1N4007 diode, featuring only a partial SCR inside the TEM lamella, an increasing capacitance (up to approx. $C = \SI{430}{\pico\farad}$) was observed, revealing a MOSCAP-like behavior, which fits well with the saturation behavior of the measured I-V characteristic (fig.~2 \cite{suppl-material}). Both results clearly demonstrate the advantages of a dynamic imaging method with high spatiotemporal resolution, through which static physical properties of samples (e.g. capacitance), remaining hidden in static experiments, can be uncovered. This clearly affirms the initial question and shows that the potential-sensitive EH, in particular, can be improved into local dynamic potentiometry by means of iGate.

While for the used extended Lorentz-mode (only used due to the large dimension of the investigated diodes) the spatial resolution is limited to approx. $\SI{30}{\nm}$ and the temporal resolution to approx. $\SI{25}{\ns}$ (due to the bandwidth limitations of the used signal generator), a transition to higher magnifications (e.g. common medium- or even high-resolution TEM modes) is inherently given by the TEM, opening the door to the investigation of smaller nanostructures (e.g. dynamic of semiconductor/magnetic nanostructures). Meanwhile, the step to higher temporal resolutions (in the sub-ns regime) have already been demonstrated using self-developed control devices \cite{Wagner2022}. In addition, iGate can in principle be applied to further interferometric techniques (e.g. inline electron holography, ptychography or diffraction), hence, creating a completely new pathway for the acquisition of movies of in-operando nanostructures.

\bibliography{references}

\end{document}



\title{Imaging Localized Variable Capacitance During Switching Processes in Silicon Diodes by Time-Resolved Electron Holography}

\author{Tolga Wagner}
 \affiliation{Technische Universit{\"a}t Berlin, Institute of Optics and Atomic Physics, Stra{\ss}e des 17. Juni 135, Berlin 10623, Germany}
 \email{tolga.wagner@physik.tu-berlin.de}

\author{H{\"u}seyin \c{C}elik}%
 \affiliation{Technische Universit{\"a}t Berlin, Institute of Optics and Atomic Physics, Stra{\ss}e des 17. Juni 135, Berlin 10623, Germany}

\author{Dirk Berger}%
 \affiliation{Technische Universit{\"a}t Berlin, Center for Electron Microscopy, Stra{\ss}e des 17. Juni 135, Berlin 10623, Germany}

\author{Ines H{\"a}usler}%
 \affiliation{Humboldt-Universit{\"a}t zu Berlin, Department of Physics, Unter den Linden 6, Berlin 12489, Germany}

\author{Michael Lehmann}%
 \affiliation{Technische Universit{\"a}t Berlin, Institute of Optics and Atomic Physics, Stra{\ss}e des 17. Juni 135, Berlin 10623, Germany}

\date{\today}

\maketitle
\section{Electrical Characterization of the UG1A and 1N4007 Diodes}

During preparation, both diodes were electrically characterized by means of $I$-$V$-characterization and switching time analyses.

\begin{figure}[h!]
    \includegraphics[width=\columnwidth]{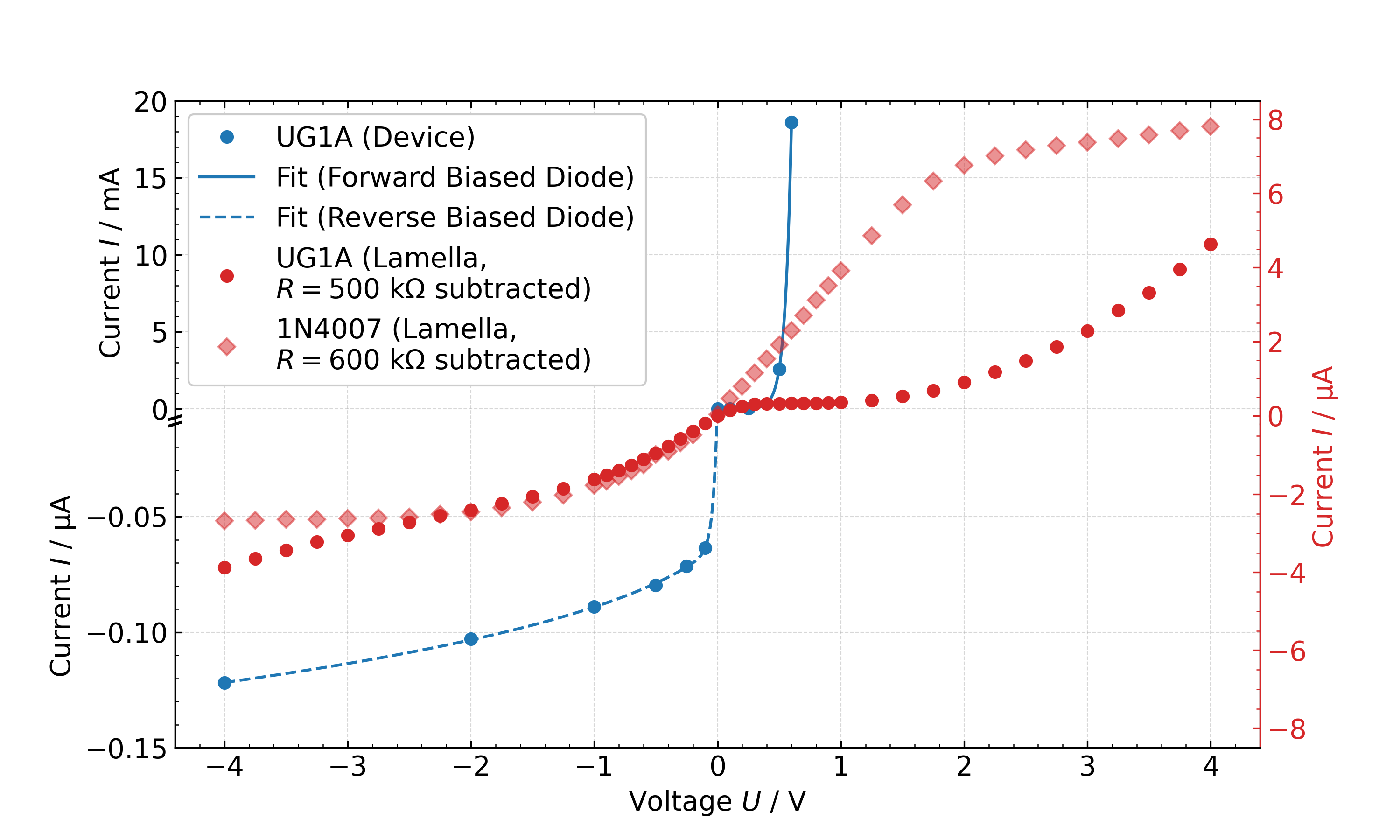}
    \caption{Plot of the current-voltage characteristic ($I$-$V$ curve) of the diodes. $I$-$V$ curve of the intact UG1A device in a voltage range from $U = \SI{-4.0}{\volt}$ to $U = \SI{0.7}{\volt}$ (blue dots) with a corresponding fit (blue line for the forward biased diode and blue dashed for the reverse biased diode). $I$-$V$ curves of the prepared TEM lamellae (shunt resistances subtracted) in a voltage range from $U = \SI{-4.0}{\volt}$ to $U = \SI{-4.0}{\volt}$ (red dots for the UG1A lamella and red diamonds for the 1N4007 lamella).}
    \label{fig:UG1A_IV-Curve}
\end{figure}

Fig.~\ref{fig:UG1A_IV-Curve} shows a plot of the $I$-$V$ curves of the unprepared macroscopic (blue dots) and the FIB prepared UG1A diode (red dots). The macroscopic component shows the typical $I$-$V$ characteristic of a diode (Shockley equation for forward and reverse biased diodes, dashed blue lines in fig.~\ref{fig:UG1A_IV-Curve}):
\begin{align}
I_{DF}\left(U\right) &= I_{SF}\left[ \exp\left(\frac{eU}{n_F k_B T}\right) -1 \right], \quad &U > 0 \label{eq:I_DF} \\
I_{DR}\left(U\right) &= I_{SR}\left[ \exp\left(\frac{eU}{n_R k_B T}\right) -1 \right] \left[\left(1 - \frac{U}{U_{diff}}\right)^2 + 0.005 \right]^{\frac{m_S}{2}}, \quad &U < 0, \label{eq:I_DR}
\end{align}
where forward and reverse direction $I_{SF}$ and $I_{SR}$ denote the (leakage) saturation currents, $n_{SF}$ and $n_{SR}$ the emission coefficients, $k_B$ the Boltzmann constant, $e$ the elementary charge, $T$ the temperature, $U_{diff}$ the diffusion voltage and $m_S$ the capacitance coefficient. The fit parameters provide saturation currents of $I_{SF} = \SI{0.134 \pm 0.005}{\uA}$ and $I_{SR} = \SI{0.062 \pm 0.015}{\uA}$ and a diffusion voltage of $U_{diff} = \SI{0.354 \pm 0.080}{\volt}$. The latter is an indication for a low doping concentration which corresponds well with the strongly extended SCR in the $\si{\um}$ range. 

The current values for the prepared UG1A TEM lamella (red dots in fig.~\ref{fig:UG1A_IV-Curve}) result after subtracting an ohmic resistance (shunt resistance), which was determined by linear regression in the voltage ranges from $U = \SI{-4.0}{\volt}$ to $U = \SI{-1.0}{\volt}$ and $U = \SI{1.0}{\volt}$ to $U = \SI{4.0}{\volt}$ of the measured current values (not shown) to $R = \SI{500}{\kilo\ohm}$. For the reverse bias range, the current shows a non-linear behavior which is comparable but cannot be described exactly by eq.~(\ref{eq:I_DR}). A possible reason for this is the strong influence of surface effects on nanostructured devices (such as TEM lamellae). For the forward bias range, the current shows a similar behavior to eq.~(\ref{eq:I_DF}). However, due a weaker current slope, significantly smaller currents are reached in the prepared diode than in the device (fig.~\ref{fig:UG1A_IV-Curve}). This can be attributed to the roughly 6~orders of magnitude smaller cross section of the prepared TEM lamella (macroscopic diode approx. $\SI{1700}{\um}$ by $\SI{1700}{\um}$ and prepared TEM lamella approx. $\SI{0.3}{\um}$ by $\SI{8.2}{\um}$). Even though for the prepared diode the forward and reverse currents show similar values, the nonlinear behavior over the whole investigated voltage range and especially the similar behavior to eq.~(\ref{eq:I_DF}) of the forward biased lamella indicate an intact specimen.

Similarly, the current values for the prepared 1N4007 TEM lamella (red diamonds in fig.~\ref{fig:UG1A_IV-Curve}) result after subtracting an ohmic resistance (shunt resistance) of $R = \SI{600}{\kilo\ohm}$. These show a different trend over the entire voltage range. Although the saturation behavior (similar to eq.~(\ref{eq:I_DR})) is apparent for negative voltage values, a similar behavior can also be seen (contrary to eq.~(\ref{eq:I_DF})) for positive voltage values (inverted). Considering that the TEM lamella only features a partial SCR, the saturation behavior could be due to the accumulation of charge carriers in the area of the electrode (fig.~4 of the main paper). In contrast to a typical p-n junction, this accumulation is stationary and would lead to an increase of the capacitance with increasing reverse bias.

\begin{figure}[h!]
    \includegraphics[width=\columnwidth]{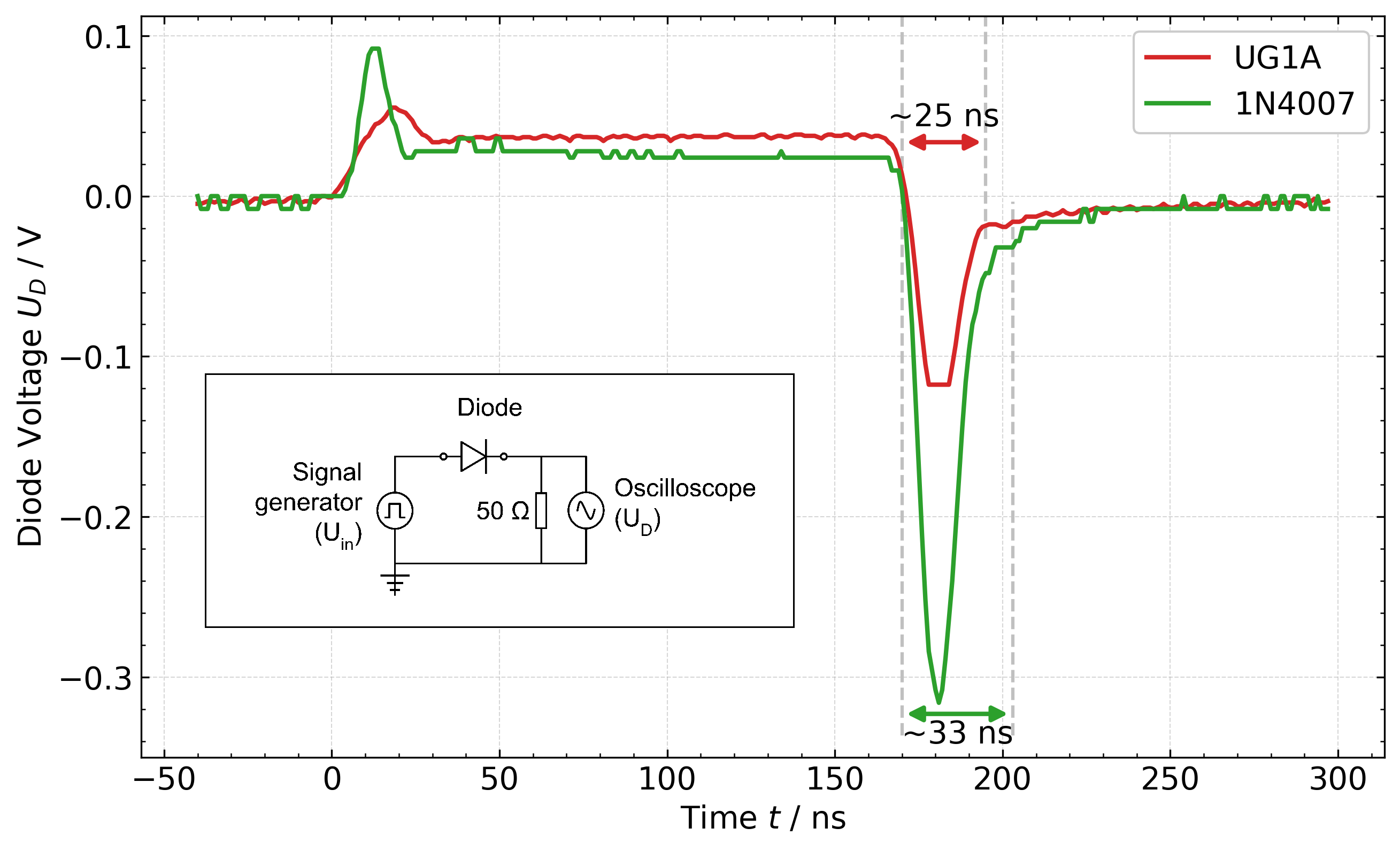}
    \caption{Plot of the diode voltage $U_D$ (measured across a $\SI{50}{\ohm}$ resistor) over the time for the determination switching time into reverse biased condition $t_{rr}$ of the diodes for small input signals. A square wave signal with a peak-to-peak amplitude of $U_{in} = \SI{1}{\volt}_{pp}$ at a repetition rate of $f = \SI{3}{\mega\Hz}$ is applied. The circuit diagram for the experimental setup is drawn in the lower left box.}
    \label{fig:ttr_UG1A_1N4007}
\end{figure}

Additionally, the switching characteristics of both macroscopic diodes were electrically characterized using the circuit shown in the lower left box in fig.~\ref{fig:ttr_UG1A_1N4007}. As an input signal, a square wave with a peak-to-peak amplitude of $U_{in} = \SI{1}{\volt}_{pp}$ at a repetition rate of $f = \SI{3}{\mega\Hz}$ was used. The switching time into reverse biased condition was determined by the diode voltage $U_D$ (measured across a $\SI{50}{\ohm}$ resistor), plotted in fig.~\ref{fig:ttr_UG1A_1N4007} (red line for UG1A, green line for 1N4007). Here, the switching times were measured to be $t_{rr} = \SI{25}{\ns}$ for the UG1A diode and $t_{rr} = \SI{33}{\ns}$ for the 1N4007 diode (indicated by the grey dashed vertical lines in fig.~\ref{fig:ttr_UG1A_1N4007}).